\documentclass[
aps,%
10pt,%
final,%
notitlepage,%
oneside,%
onecolumn,%
nobibnotes,%
nofootinbib,
superscriptaddress,%
noshowpacs,%
centertags]%
{revtex4}
\usepackage[cp1251]{inputenc}
\usepackage[T2A]{fontenc}
\usepackage[english,russian]{babel}
\usepackage{amsfonts}\usepackage{amsbsy}
\usepackage{color}
\usepackage{amsfonts}
\usepackage{amsbsy}
\usepackage{mathrsfs}
\usepackage{graphicx}
\usepackage{subfigure}
\newcommand{\eeq}{\end{equation}}
\newcommand{\bea}{\begin{eqnarray}}
\newcommand{\eea}{\end{eqnarray}}

\def\lsim{\mathrel{\rlap{
\lower4pt\hbox{\hskip-3pt$\sim$}}
    \raise1pt\hbox{$<$}}}     
\def\gsim{\mathrel{\rlap{
\lower4pt\hbox{\hskip-3pt$\sim$}}
    \raise1pt\hbox{$>$}}}     

\begin{document}





\title{Quantum simulations of thermodynamic properties of strongly coupled quark-gluon plasma 
  \footnote{Dedicated to
  A.B. Migdal on the occasion of his 100th birthday.}}%
\author{V.S.~Filinov}
\thanks{Corresponding author\quad E-mail:~\textsf{vladimir\_filinov@mail.ru}}
\affiliation{Joint Institute for High Temperatures, Russian Academy of
Sciences, Moscow, Russia}
\author{Yu.B. Ivanov}
\affiliation{GSI Helmholtzzentrum
 f\"ur Schwerionenforschung, 
Darmstadt, Germany}
\affiliation{Kurchatov Institute,
Moscow, Russia}
\author{M. Bonitz}
\affiliation{Institute for Theoretical Physics and Astrophysics, Christian Albrechts University, 
Kiel, Germany}
\author{P.R. Levashov}
\affiliation{Joint Institute for High Temperatures, Russian Academy of
Sciences, Moscow, Russia}
\author{V.E. Fortov}
\affiliation{Joint Institute for High Temperatures, Russian Academy of
Sciences, Moscow, Russia}
%



\begin{abstract}
A strongly coupled quark-gluon plasma (QGP) of heavy constituent quasi-particles
is studied by a path-integral Monte-Carlo method. 
This approach is a quantum generalization of the model developed by 
Gelman, Shuryak and Zahed. 
It is shown that this method is
able to reproduce the QCD lattice equation of state and also
yields valuable insight into the internal structure of the QGP.
The results indicate that the QGP reveals liquid-like rather than gas-like properties.
At temperatures just above the critical one it was found that
bound quark-antiquark states still survive. These states are bound by effective string-like forces 
and turns out to be colorless. At the temperature as large as twice the critical one no bound states 
are observed. Quantum effects turned out to be of prime importance in these  simulations. 
\end{abstract}

\maketitle

\section{Introduction}\label{s:intro}

Contributions of A.B. Migdal to physics of the last century  are difficult to overestimate.
These include the Landau-Migdal-Pomeranchuk effect, theory of finite Fermi systems,
pionic degrees of freedom in nuclei, physics of strong fields, {\it etc}.  This results entered many
textbooks and found numerous applications. A.B. Migdal was one of the first who started to think
about possible existence of an anomalous nuclear matter \cite{Migdal}. Nowadays this concept is, in particular,
associated with the phase transition into the QGP. While the possibility of pion
condensation is still vividly discussed in astrophysical applications, see, e.g., \cite{Voskresensky}.

 Investigation of properties of the QGP is
one of the main challenges of strong-interaction physics, both
theoretically
and experimentally.
Many features of this matter were
experimentally discovered at the Relativistic Heavy Ion Collider
(RHIC) at Brookhaven. The most striking result, obtained from analysis
of these experimental data \cite{shuryak08}, is that the deconfined
quark-gluon matter behaves as an almost perfect fluid rather than a perfect gas,
as it could be expected from the asymptotic freedom.

There are  various approaches to studying QGP.
Each approach has its advantages and  disadvantages.
The most fundamental way to compute
properties of the strongly interacting matter is provided by the lattice QCD \cite{Lattice09,Fodor09,Csikor:2004ik}.
Interpretation of these very complicated computations
requires application of various QCD motivated, albeit schematic, models simulating various aspects of the full theory.
Moreover, such models are needed in cases when the lattice QCD fails, e.g. at large
baryon chemical potentials and out of equilibrium.
While some progress has been achieved in the recent years,
we are still far away from having a satisfactory understanding of the QGP dynamics.

A semi-classical approximation, based on a point like quasi-particle picture has been introduced in \cite{LM105}.
It is expected that the main features of non-Abelian
plasmas can be understood in simple semi-classical terms without
the difficulties inherent to a full quantum field theoretical analysis.
Independently the same ideas were implemented in terms of molecular dynamics (MD) \cite{Bleicher99}.
Recently this MD approach was further developed in a series of works
\cite{shuryak1,Zahed}. The MD allowed one to treat soft processes in the QGP which
are not accessible by perturbative means.


A strongly correlated behavior of the QGP is expected to show up in long-ranged spatial correlations of quarks and
gluons which, in fact, may give rise to liquid-like and, possibly, solid-like structures.
This expectation is based on a very similar
behavior observed in electrodynamic plasmas  \cite{shuryak1,thoma04,afilinov_jpa03}. This similarity was exploited
to formulate a classical non-relativistic model of a color Coulomb interacting QGP \cite{shuryak1} which was  numerically analyzed by classical MD simulations.
Quantum effects were either neglected or included
phenomenologically via a short-range repulsive correction to the pair potential. Such a rough model may become
a critical issue at high densities,
where quantum effects strongly affects properties of the QGP.
Similar models have been used in
electrodynamic plasmas and showed poor behavior in the region of strong wave function overlap, in particular at the Mott density.
For temperatures and densities of the QGP considered
in  Ref. \cite{shuryak1} these effects are very important as the quasi-particle
thermal wave length is of order the average interparticle distance.


In this paper we
extend previous classical nonrelativistic simulations \cite{shuryak1}
based on a color Coulomb interaction to the quantum regime.
We develop an  approach based on path integral Monte Carlo (PIMC)
simulations of the strongly coupled QGP which
self-consistently takes  into account the Fermi (Bose) statistics of quarks (gluons).
Following an idea of Kelbg \cite{kelbg}, quantum corrections to the pair potential
are rigorously derived \cite{dusling09}.
To extend the method of
quantum potentials to a stronger coupling, an ``improved Kelbg potential''
was derived, which contains a single free parameter,
being fitted to the exact solution of the quantum-mechanical two-body problem.
Thus, the method of the improved Kelbg potential is able to describe
thermodynamic properties up to moderate couplings \cite{afilinov_pre04}.
However, this approach may fail, if bound states of more than two
particles are formed in the system. This results in
a break-down of the pair approximation for the density matrix, as demonstrated in Ref. \cite{afilinov_pre04}.
A superior approach, which does not have this limitation, consists in use the original Kelbg potential in the PIMC
simulations which effectively map the problem onto a high-temperature weakly coupled and weakly degenerate one.
This allows one to rigorously extend the analysis to strong couplings and is, therefore,
a relevant choice for the present purpose.

This method has been successfully applied to strongly coupled electrodynamic plasmas
\cite{filinov_ppcf01,bonitz_jpa03}. 
Examples are partially-ionized dense hydrogen plasmas, where liquid-like and
crystalline behavior was observed \cite{filinov_jetpl00,bonitz_prl05}.
Moreover, also partial ionization effects and pressure ionization
could be studied from first principles \cite{filinov_jpa03}.
The same methods have been also applied to electron-hole plasmas in
semiconductors \cite{bonitz_jpa06,filinov_pre07}, including excitonic bound states,
which have many similarities to the QGP due to
smaller mass differences as compared to electron-ion plasmas.

The main goal of this article is to test the developed approach for ability to reproduce known lattice data \cite{Lattice09,Fodor09} and to predict other properties of the QGP, which are still unavailable for the
lattice calculations.
To this end we use a simple model \cite{shuryak1} of the QGP consisting of quarks, antiquarks and gluons interacting via a color Coulomb potential.
First results of applications of the PIMC method to study of thermodynamic properties of the nonideal QGP
have already been briefly reported in \cite{Filinov:2009pimc,Filinov:2010pimc}.
In this paper we present a comprehensive report on the
thermodynamic properties.

\section{Thermodynamics of QGP} \label{Thermodynamics}
\subsection{Basics of the model}\label{semi:model}

Our model is based on a resummation technique and lattice simulations for 
dressed quarks, antiquarks and gluons interacting via the color Coulomb potential.
The assumptions of the model are similar to those of Ref. \cite{shuryak1}:
%
\begin{description}
 \item[I.]
 All color quasi-particles are heavy, i.e. their mass ($m$) is higher than the mean kinetic energy
 per particle. For instance, at zero net-baryon density it amounts to $m > T$, where $T$ is a temperature.
 Therefore these particles move non-relativistically. This assumption is based on the analysis of lattice data \cite{Lattice02,LiaoShuryak}.
 \item[II.] Since the particles are non-relativistic, interparticle interaction is dominated
 by a color-electric Coulomb potential, see Eq. (\ref{Coulomb}).
 Magnetic effects are neglected as sub-leading ones. 
 \item[III.] Relying on the fact that the color representations are large, the color operators 
 are substituted by their average values, i.e. by classical color vectors, the
time evolution of which 
is described
by Wong's dynamics \cite{Wong}.
\end{description}
The quality of these approximations and their limitations were discussed in Ref. \cite{shuryak1}.
Thus, 
this model requires the following quantities as an input:
\begin{description}
\item[1.] the quasi-particle mass, $m$, and
\item[2.] the coupling constant $g^2$.
\end{description}
All the input quantities should be deduced from the lattice data or from an appropriate model simulating these data.

\subsection{Path-Integral Monte-Carlo Simulations}\label{s:pimc}

Thus,  we consider a three-component QGP consisting of a number of dressed quarks ($N_q$),
antiquarks $(N_{\bar{q}})$ and   gluons $(N_g)$  represented by quasi-particles.
 In thermal equilibrium
 the average values of these numbers can be found in the grand canonical ensemble defined by  the  temperature-dependent Hamiltonian,
which can be written as ${\hat{H}}={\hat{K}}+{\hat{U}}$. The kinetic and color Coulomb interaction energy of the quasi-particles are
\begin{eqnarray}
\label{Coulomb}
{\hat{K}}=
\sum_i \left[m_i(T,\mu_q)+
\frac{\hat{p}^2_i}{2m_i(T,\mu_q)}\right],
\qquad
{\hat{U}_C}=\frac{1}{2}\sum_{i,j}
\frac{g^2(|r_i-r_j|,T,\mu_q)
\langle Q_i|Q_j \rangle}{4\pi|r_i-r_j|},
\end{eqnarray}
Here the $Q_i$ denote the Wong's color variable
(8-vector in the $SU(3)$ group), 
$T$ is the temperature and $\mu_q$ is
the quark chemical potential, 
$\langle Q_i|Q_j \rangle$ denote scalar product of color vectors. 
In fact, the quasi-particle mass and the coupling constant,
as deduced from the lattice data, are functions of $T$ and, in general, $\mu_q$.
Moreover, $g^2$ is a function of distance $r$, which produces a linearly rising potential
at large $r$ \cite{Rich}.

The thermodynamic properties in the grand canonical ensemble with given temperature $T$,
chemical potential $\mu_q$ and fixed volume $V$ are fully described by the
grand partition function
\begin{eqnarray}\label{Gq-def}
&&Z\left(\mu_q,\beta,V\right)=
\sum_{N_q,N_{\bar{q}},N_g}\frac{\exp(\mu_q(N_q-N_{\bar{q}})/T)}{N_q!N_{ \bar{q}}!N_g!} \sum_{\sigma}\int\limits_V
dr dQ \,\rho(r,Q, \sigma ; N_q,N_{\bar{q}},N_g;\beta),
\end{eqnarray}
%
%
%
where $\rho(r,Q, \sigma ; N_q,N_{\bar{q}},N_g;\beta)$ denotes the diagonal matrix
elements of the density operator ${\hat \rho} = \exp (- \beta{\hat H})$, and $\beta=1/T$.
Here $\sigma$, $r$ and $Q$
denote the spin, spatial and color degrees of freedom 
of all quarks, antiquarks and gluons in the ensemble, respectively.
Correspondingly, the $\sigma$ summation and integration $dr dQ$ run over
all individual degrees of freedom of the particles.
%
Since the masses and the coupling constant depend on the temperature and chemical potential,
special care should be taken to preserve thermodynamical consistency of this approach.
In order to preserve the thermodynamical consistency,
thermodynamic functions such as pressure, $P$, entropy, $S$, baryon number, $N_B$, and
internal energy, $E$, should be calculated through respective derivatives of
 the logarithm of the partition function
\begin{eqnarray}
\label{p_gen}
P=\partial (T\ln Z) / \partial V, \quad
S=\partial (T\ln Z) / \partial T, \quad
N_B=(1/3)\partial (T\ln Z) / \partial \mu_q, \quad
E= -PV+TS+3 \mu_q N_B.
\end{eqnarray}
This is a conventional way of maintaining the thermodynamical consistency in approaches
of the Ginzburg-Landau type as they are applied in high-energy physics.

The exact density matrix of interacting quantum
systems can be constructed using a path integral
approach~\cite{feynm,zamalin}
based on the operator identity
$e^{-\beta {\hat H}}= e^{-\Delta \beta {\hat H}}\cdot
e^{-\Delta \beta {\hat H}}\dots  e^{-\Delta \beta {\hat H}}$,
where the r.h.s. contains $n+1$ identical factors with $\Delta \beta = \beta/(n+1)$.
which allows us to
rewrite\footnote{For the sake of notation convenience, we ascribe superscript $^{(0)}$
to the original variables.}
the integral in Eq.~(\ref{Gq-def})
\begin{eqnarray}
&&
\sum_{\sigma} \int\limits dr^{(0)}dQ^{(0)}\,
\rho(q^{(0)},Q^{(0)},\sigma; N_q,N_{\bar{q}},N_g;\beta) =
\int\limits  dr^{(0)}dQ^{(0)} dr^{(1)}dQ^{(1)}\dots
dr^{(n)}dQ^{(n)} \, \rho^{(1)}\cdot\rho^{(2)} \, \dots \rho^{(n)}
\nonumber\\&\times&
\sum_{\sigma}\sum_{P_q} \sum_{P_{ \bar{q}}}\sum_{P_g}(- 1)^{\kappa_{P_q}+ \kappa_{P_{\bar{q}}}} \,
{\cal S}(\sigma, {\hat P_q}{\hat P_{ \bar{q}}}{\hat P_g} \sigma^\prime)\,
{\hat P_q} {\hat P_{ \bar{q}}}{\hat P_g}\rho^{(n+1)}\big|_{r^{(n+1)}= r^{(0)}, \sigma'=\sigma} \,
\nonumber\\&=&
\int\limits dQ^{(0)}dr^{(0)} dr^{(1)}\dots dr^{(n)}
\tilde{\rho}(r^{(0)},r^{(1)}, \dots r^{(n)};Q^{(0)}; N_q,N_{\bar{q}},N_g;\beta).
 \label{Grho-pimc}
\end{eqnarray}
The spin gives rise to the spin part of the density matrix (${\cal
S}$) with exchange effects accounted for by the permutation
operators  $\hat P_q$, $\hat P_{ \bar{q}}$ and $\hat P_g$ acting on the quark, antiquark and gluon spatial $r^{(n+1)}$
and color $Q^{(n+1)}$ coordinates,
as well as on the spin projections $\sigma'$. The
sum runs over all permutations with parity $\kappa_{P_q}$ and
$\kappa_{P_{ \bar{q}}}$. In Eq.~(\ref{Grho-pimc}) the index $l=1\dots n+1$
labels the off-diagonal
density matrices
$\rho^{(l)}\equiv \rho\left(r^{(l-1)},Q^{(l-1)};r^{(l)},Q^{(l)};\Delta\beta\right) \approx
\langle r^{(l-1)}|e^{-\Delta \beta {\hat H}}|r^{(l)}\rangle\delta_\epsilon(Q^{(l-1)}-Q^{(l)})$, where
$\delta_\epsilon(Q^{(l-1)}-Q^{(l)})$ is a delta-function at $\epsilon\rightarrow 0$.
Accordingly each $a$ particle is represented by a set of $n+1$ coordinates
(``beads''),
i.e. by $(n+1)$ 3-dimensional vectors
$\{r_a^{(0)}, \dots r_a^{(n)}\}$
and a 8-dimensional color
vector $Q^{(0)}$ in the $SU(3)$ group, since all beads
are characterized by the same color charge.

The main advantage of decomposition (\ref{Grho-pimc}) is that it
allows us to use a  semi-classical  approximation for density matrices $\rho^{(l)}$,
which is applicable due to smallness of artificially introduced factor $1/(n+1)$.
This parameter makes the thermal  wavelength $\Delta\lambda_a=\sqrt{2 \pi \Delta\beta/m_a}$
of a bead of type $a$ ($a = q, \overline{q}, g$),
smaller then a characteristic scale
of variation of the potential energy.
In the high-temperature limit $\rho^{l}$ can be approximated by a product of
two-particle density matrices.
Generalizing the
electrodynamic plasma results \cite{filinov_ppcf01} to the case of an additional bosonic species (i.e. gluons),
we write
\begin{eqnarray}
&&
\tilde{\rho}(r^{(0)},r^{(1)}, \dots r^{(n)};Q^{(0)}; N_q,N_{\bar{q}},N_g;\beta)
\nonumber\\&=&
 \sum_{s,k}\frac{C^s_{N_q}}{2^{N_q}} \frac{C^k_{N_{ \bar{q}}}}{2^{N_{\bar{q}}}}
\frac{\exp\{-\beta U(r,Q,\beta)\}}{\lambda_q^{3N_q} \lambda_{{ \bar{q}}}^{3N_{ \bar{q}}}\lambda_g^{3N_g}}\,
 \, {\rm per}\,||\tilde{\phi}^{n,0}||_{\rm glue} \,
{\rm det}\,||\tilde{\phi}^{n,0}||_s \, {\rm det}\,||\tilde{\phi}^{n,0}||_k \,
\prod\limits_{l=1}^n \prod\limits_{p=1}^N
\phi^l_{pp} \, \label{Grho_s}
\end{eqnarray}
where $N=N_q+N_{\bar{q}}+N_g$, $s$ and $k$ are numbers of quarks  and antiquarks, respectively,
with the same spin projection, $\lambda_a=\sqrt{2 \pi \beta / m_a}$
 is a thermal  wavelength of an $a$ particle,
$C^s_{N_a}=N_a!/[s!(N_a-s)!]$, the antisymmetrization and
symmetrization are taken into account by the symbols ``det'' and ``per'' denoting the determinant and permanent, respectively.
Functions
$\phi^l_{pp}\equiv \exp\left[-\pi\left|\xi^{(l)}_p\right|^2\right]$ and matrix elements
$\tilde{\phi}_{to}^{n,0}=\exp \left(-\pi
\left|(r_{t}^{(0)}-r_{o}^{(0)})+ y_{t}^{(n)}\right|^2/\Delta\lambda_{a}^2\right)
\delta_\epsilon(Q_t-Q_o)$, where $t$ and $o$ are particle's indexes, are expressed in
terms of
distances ($y_{a}^{(1)}, \dots , y_{a}^{(n)}$) and
dimensionless distances ($\xi_{a}^{(1)}, \dots , \xi_{a}^{(n)}$) between
neighboring beads of an $a$ particle, defined as
$r_{a}^{(l)} = r_{a}^{(0)}+y_{a}^{(l)}$, ($l>0$),
and $y_a^{(l)}=\Delta\lambda_a\sum_{k=1}^{l}\xi_a^{(k)}$.
Notice that the indices $s$ and $k$ in $\rm det\,||\tilde{\phi}^{n,0}||_s$ and  
$\rm det\,||\tilde{\phi}^{n,0}||_k$  denoted that matrices $||\tilde{\phi}^{n,0}||$ have two nonzero blocks 
related to quark and antiquark quasi-particles with the same spin projections. 
%
The density matrix (\ref{Grho_s}) has been transformed to a form which does not
contain an explicit sum over permutations
 and thus no sum of terms with alternating sign (in the case of quarks and antiquarks).
Let us stress that the determinants depend also on the color variables.

In Eq.~(\ref{Grho_s}) the total color interaction energy
\begin{eqnarray}
U(r,Q,\beta) = \frac{1}{n+1}\sum_{l=1}^{n+1}\tilde{U}^{(l) }=
\frac{1}{n+1}\sum_{l=1}^{n+1} \frac{1}{2}
\sum_{p\neq t}\Phi^{pt}(|r_p^{(l-1)}-r_t^{(l-1)}|,|r_p^{(l)}-r_t^{(l)}|, Q_p,Q_t)
\label{up}
\end{eqnarray}
is defined in terms of
off-diagonal two-particle effective quantum potential
$\Phi^{pt}$,
which is obtained by expanding the two-particle density matrix $\rho_{pt}$
up to the first order in small parameter $1/(n+1)$:
\begin{eqnarray}
&&
\rho_{pt}(r,r',Q_p,Q_t,\Delta\beta) \approx \rho_{pt}^0(r,r',Q_p,Q_t,\Delta\beta)-
\int_0^1 d\tau \int dr'' 
\frac{\Delta\beta g^2(|r''|,T,\mu_q)\langle Q_p|Q_t \rangle}{4\pi|r''|\Delta\lambda_{pt}^2\sqrt{\tau(1-\tau)}}
\,\nonumber\\&\times&
\exp\left(-\frac{\pi|r'-r''|^2}{\Delta\lambda_{pt}^2(1-\tau)}\right)
\exp\left(-\frac{\pi|r''-r|^2}{\Delta\lambda_{pt}^2\tau}\right)
\approx \rho_{pt}^0
\exp[-\Delta\beta
\Phi^{pt}(r,r', Q_p,Q_t)].
\label{GPERT}
\end{eqnarray}
where $ r= r_p-r_t$, $r'= r'_p-r_t'$,
%
$\Delta\lambda_{pt}=\sqrt{2\pi\Delta\beta /m_{pt}},$
%
$m_{pt}=m_{p}m_{t}/(m_{p}+m_{t})$
is the reduced mass of the $(pt)$-pair of particles,
 and $\rho_{pt}^0$ is the  two-particle
density matrix of the ideal gas.
The result for the diagonal color Kelbg potential can be written as
\begin{eqnarray}
\Phi^{pt}( r,r,Q_p,Q_t) 
\approx \frac{g^2(T,\mu_q)\,\langle Q_p|Q_t \rangle}{4 \pi \Delta\lambda_{pt} x_{pt}} \,\left\{1-e^{-x_{pt}^2} +
\sqrt{\pi} x_{pt} \left[1-{\rm erf}(x_{pt})\right] \right\},
\label{kelbg-d}
\end{eqnarray}
where $x_{pt}=| r_{p}-r_{t}|/\Delta\lambda_{pt}$.
Here the function $g^2(T, \mu_q) = \overline{g^2(r'',T. \mu_q)}$,
resulting from averaging of the initial
$g^2(r'', T, \mu_q)$ over relevant distances of order $\Delta\lambda_{pt}$,
plays the role of an effective coupling constant.
Note that the color Kelbg
potential approaches the color Coulomb potential
at distances larger than $\Delta\lambda_{pt}$. What is of prime importance, the color Kelbg
potential is finite at zero distance, thus removing
in a natural way the classical divergences and making any artificial cut-offs obsolete.
This potential 
is a straightforward generalization of the corresponding potential of electrodynamic plasmas
\cite{afilinov_pre04}. 
The off-diagonal 
elements of the effective interaction are approximated
 by the diagonal one by means of
$\Phi^{pt}(r,r';,Q_p,Q_t)\approx [\Phi^{pt}(r,r,Q_p,Q_t) + \Phi^{pt}(r',r',Q_p,Q_t)]/2$.

The described path-integral representation of the density matrix
is exact in the limit $n\to \infty$. For any finite
number $n$, the error of the above approximations for the whole product on the r.h.s. of Eq.
(\ref{Grho-pimc}) is of the order $1/(n+1)$ whereas the error of each $\rho^{l}$ is
of the order $1/(n+1)^2$, as it was shown in Ref. \cite{filinov_ppcf01}.

The main contribution to the partition function comes from
configurations in which the `size' of the cloud of beads of quasi-particles is of
the order of their thermal  wavelength
$\lambda_a$
whereas characteristic
distances  between beads of each quasi-particle are of the order of
$\Delta\lambda_a$.

\section{Simulations of QGP}\label{s:model}

To test the developed approach we consider the QGP 
only at zero baryon density 
and further
simplify the model by additional approximations, similarly to Ref. \cite{shuryak1}:
\begin{description}
\item[IV] We replace the grand canonical ensemble by a canonical one.
The thermodynamic properties in the canonical ensemble with given temperature $T$ and fixed volume $V$ are fully
described by the density operator ${\hat \rho} = e^{-\beta {\hat H}}$ with the partition function
defined as follows
\begin{equation}\label{q-def}
Z(N_q,N_{ \bar{q}},N_g,V;\beta) = \frac{1}{N_q!N_{ \bar{q}}!N_g!} \sum_{\sigma}\int\limits_V
dr dQ\,\rho(r,Q, \sigma ;\beta),
\end{equation}
with $N_q=N_{ \bar{q}}$ and hence $N_B=0$.
In order to preserve the thermodynamical consistency of this formulation,
thermodynamic quantities
should be calculated through respective derivatives of
the logarithm of the partition function similarly to that in Eq. (\ref{p_gen})
with the exception that now $N_a$ are independent variables.

\item[V] Since the masses of quarks of different flavors extracted from lattice data are very similar,
we do not distinguish between quark flavors.
Moreover, we take the quark and gluon quasi-particle masses being equal because their values
deduced from the lattice data \cite{Lattice02,LiaoShuryak} are very close.
\item[VI]  Because of the equality of masses and approximate equality of number of degrees of freedom
of quarks, antiquarks and gluons, we assume that these species
are equally
represented in the system: $N_q = N_{\bar{q}} = N_g$.
\item[VII] For the sake of technical simplicity, the $SU(3)$ color group is replaced by $SU(2)$.
\end{description}
Thus, this simplified model requires an additional quantity as an input:
\begin{description}
\item[3.] the density of quasi-particles $(N_q+N_{\bar{q}}+N_g)/V=n(T)$ as a function of the temperature.
\end{description}
Although this density is unknown from the QCD lattice calculations and  we use it as a fit
parameter, it is very important to partially overcome constrains of the above simplifications.
First, it concerns the use of the $SU(2)$ color group, which first of all reduces the degeneracy factors of the
quark and gluon states, as compared to the $SU(3)$ case, and thereby reduces pressure and all other thermodynamic
quantities. A proper fit of the density allows us to remedy this deficiency of the normalization.
Second, in fact we consider the system of single quark flavor, i.e. all quarks are identical,
which also reduces the normalization of all thermodynamic quantities. The density fit cures
the deficiency of this normalization, though the excessive anticorrelation of quarks remains.

\begin{figure}[htb]
\vspace{0cm} \hspace{0.0cm}
\includegraphics[width=7.9cm,clip=true]{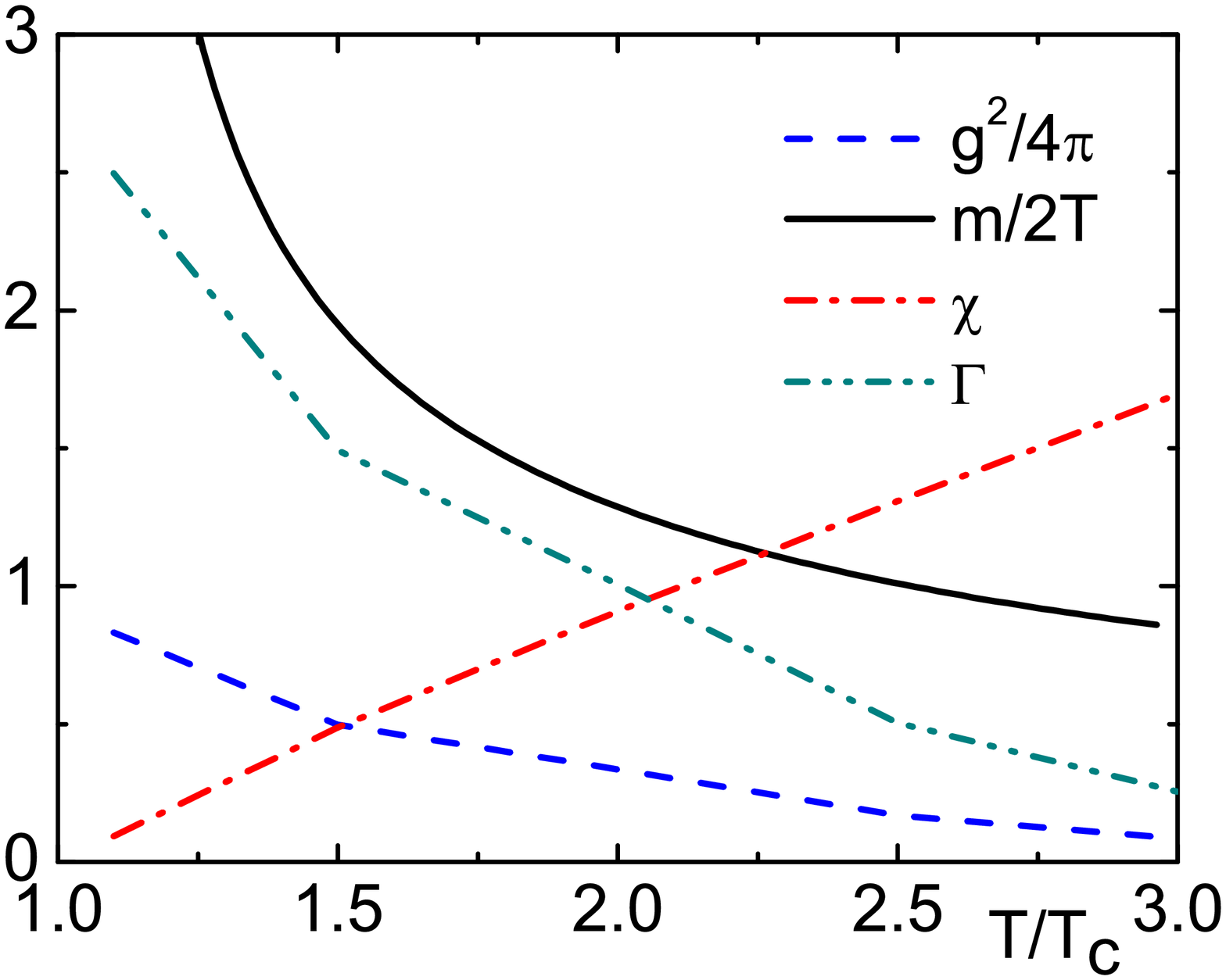}
\includegraphics[width=7.9cm,clip=true]{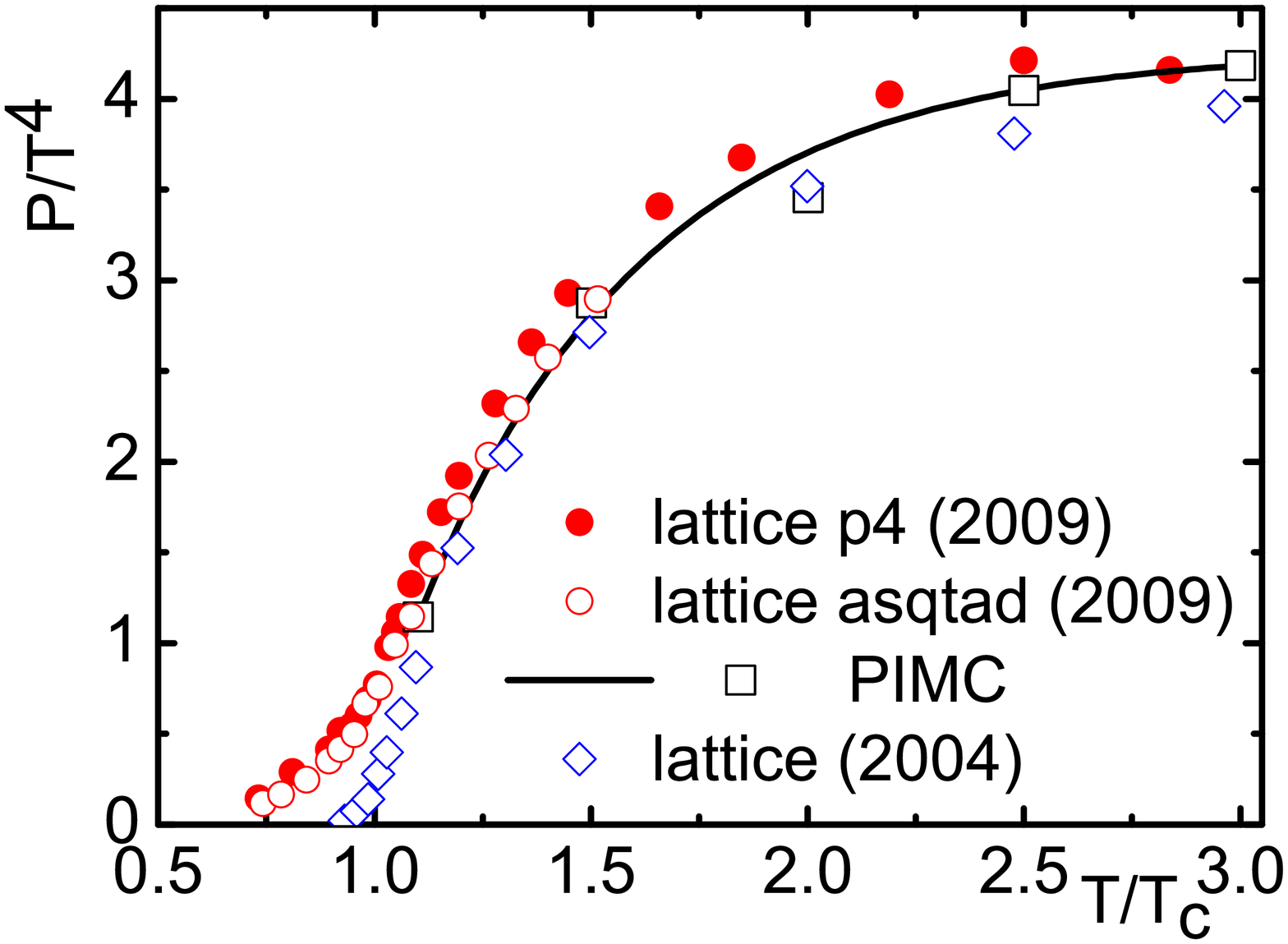}
\caption{Left panel: Temperature dependence of the model input
  quantities, coupling constant $g^2$ and mass-to-temperature ratio
  (scaled by 1/2), 
the plasma coupling parameter $\Gamma$ [see Eq. (\ref{Gamma})] and the
degeneracy parameter $\chi$. The $\chi$ parameters for different
species are equal, since their masses and densities are assumed to be
equal. 
Right panel:
Equation of state (pressure versus temperature) of the QGP from PIMC
simulations (open squares) compared 
to lattice data of Refs.~\cite{Lattice09,Csikor:2004ik}.
The solid line is a smooth interpolation between the PIMC points. 
Results of the HotQCD Collaboration \cite{Lattice09} are presented by
filled circles, while results of the Budapest group
\cite{Csikor:2004ik}, open circles. Different kinds of circles
(filled and open) correspond to different discretization schemes of the
QCD action (p4 and asqtad, see \cite{Lattice09} for details). 
}
\label{fig:EOS}
\end {figure}

Ideally the parameters of the model should be deduced from the QCD lattice data. However, presently
this task is still quite ambiguous. Therefore, in the present simulations we take a possible (maybe,
not the most reliable) set of parameters.
Following Refs. \cite{LiaoShuryak,shuryak1}, the parametrization
of the quasi-particle mass is taken in the form
\begin{eqnarray}
\label{mass}
m(T)/T_c=0.9/(T/T_c-1)+3.45+0.4T/T_c
\end{eqnarray}
where $T_c=175$ MeV is the critical temperature. This parametrization fits the quark mass at two values
of temperature obtained in the lattice calculations \cite{Lattice02}. According to \cite{Lattice02} the masses
are quite large: $m_q/T \simeq 4$  and $m_g/T \simeq 3.5$. These are essentially larger than masses
required for quasi-particle fits \cite{Peshier96,Ivanov05} of the lattice  thermodynamic properties
of the QGP:  $m_q/T \simeq 1\div 2$  and $m_g/T \simeq 1.5\div 3$.
Moreover, the pole quark mass $m_q/T \simeq 0.8$ was reported
in recent work \cite{Karsch09a}, as deduced from lattice calculations. Nevertheless, in spite of the fact
that it obviously produces too high masses, we use parametrization
(\ref{mass}) in order to be compatible with the input of classical MD
of Ref.  \cite{shuryak1}.
The $T$-dependence of this mass is illustrated in Fig.~\ref{fig:EOS}
(left panel). 

The coupling constant, i.e. $\alpha_s = g^2/(4\pi)$, used in the
simulations is displayed in the left panel of 
Fig.~\ref{fig:EOS} as well. As seen, $\alpha_s$
well complies with phenomenologic QCD estimations \cite{Prosperi} 
of its values. Notice that in previous publications
\cite{Filinov:2009pimc,Filinov:2010pimc} the factor $(N_c^2-1)$, where
$N_c$ is the number of colors in the $SU(N_c)$ color group, was
accidentally included in $g^2$, when displaying it 
corresponding figures. In fact, this factor is a part of Casimirs
defining the normalization of the color vectors in the color group.

The density of quasi-particles, which is additionally required within the canonical-ensemble
approach, was chosen on the condition of the best agreement of the calculated pressure with the corresponding
lattice result, see Fig.~\ref{fig:EOS} (right panel). It was taken to be $n(T)=0.24 T^3$. From the first
glance, it is a very low density. For example, in the classical simulations of Ref. \cite{shuryak1} it was taken
as $n(T)/T^3=6.3$, which corresponds to the density of an ideal gas of {\em massless} quarks, antiquarks
and gluons. Since the quasi-particles are very heavy in the present model (as well as in that of
Ref. \cite{shuryak1}), 
the latter density looks unrealistically high. Even in
quasi-particle models \cite{Peshier96,Ivanov05}, where the masses are lower,
the density turns out to be $n(T)/T^3 \approx 1.4$.
Since Eq. (\ref{mass}) gives even
larger masses than those in Refs. \cite{Peshier96,Ivanov05} and in view of the adopted large coupling,
the chosen value of $n(T)$ does not
look too unrealistic. 

Thus, although the chosen set of parameters is still debatable, it is somehow self-consistent.
In the future we are going to get rid of the $n(T)$ parameter, by applying the grand-canonical approach,
and by using more moderate (and maybe realistic) sets of parameters.

Calculation of the equation of state (right panel of Fig.~\ref{fig:EOS})
was used to optimize the parameters of the model in order to
proceed to predictions of other properties concerning the internal structure
and in the future also non-equilibrium dynamics of the QGP.
The plasma coupling parameter is defined as 
\begin{eqnarray}
\label{Gamma}
\Gamma = \frac{ \overline{q_2} g^2}{4\pi r_s T}
\end{eqnarray}
where  $r_s$ is the the Wigner-Seitz radius, defined such that
$4\pi r_s^3/3 = n $, and $\overline{q_2}$ the quadratic Casimir value
averaged over quarks, antiquarks and gluons,
$\overline{q_2}=N_c^2-1$ is a good estimate for this quantity. 
The plasma parameter is a measure of 
ratio of the average potential to the average kinetic energy. It 
is also presented in
Fig.~\ref{fig:EOS} (left panel). It turns out to be of the order of
unity which indicates 
that the QGP is a strongly coupled Coulomb liquid rather than a gas.
In the studied temperature range, $1<T/T_c<3$, the QGP  is, in fact,
quantum degenerate, since 
the degeneracy parameter
$\chi_a = n_a\lambda_a^3$
(where the thermal wave length, $\lambda$, is  defined in the
previous sect.)
varies from $0.1$ to $1.7$, see Fig.~\ref{fig:EOS} (left panel).

\begin{figure}[htb]
\vspace{0cm} \hspace{0.0cm}
\includegraphics[width=7.9cm,clip=true]{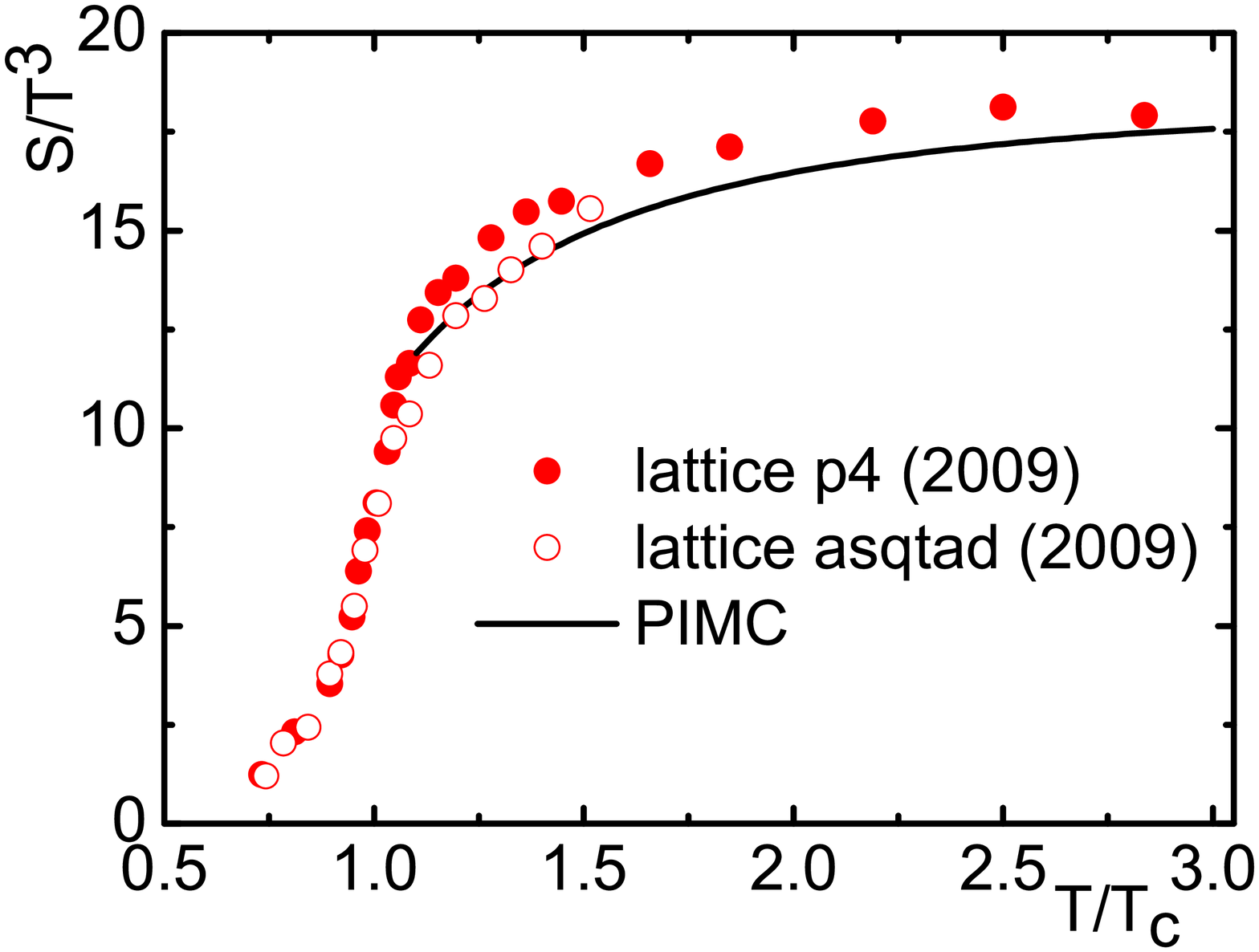}
\includegraphics[width=7.9cm,clip=true]{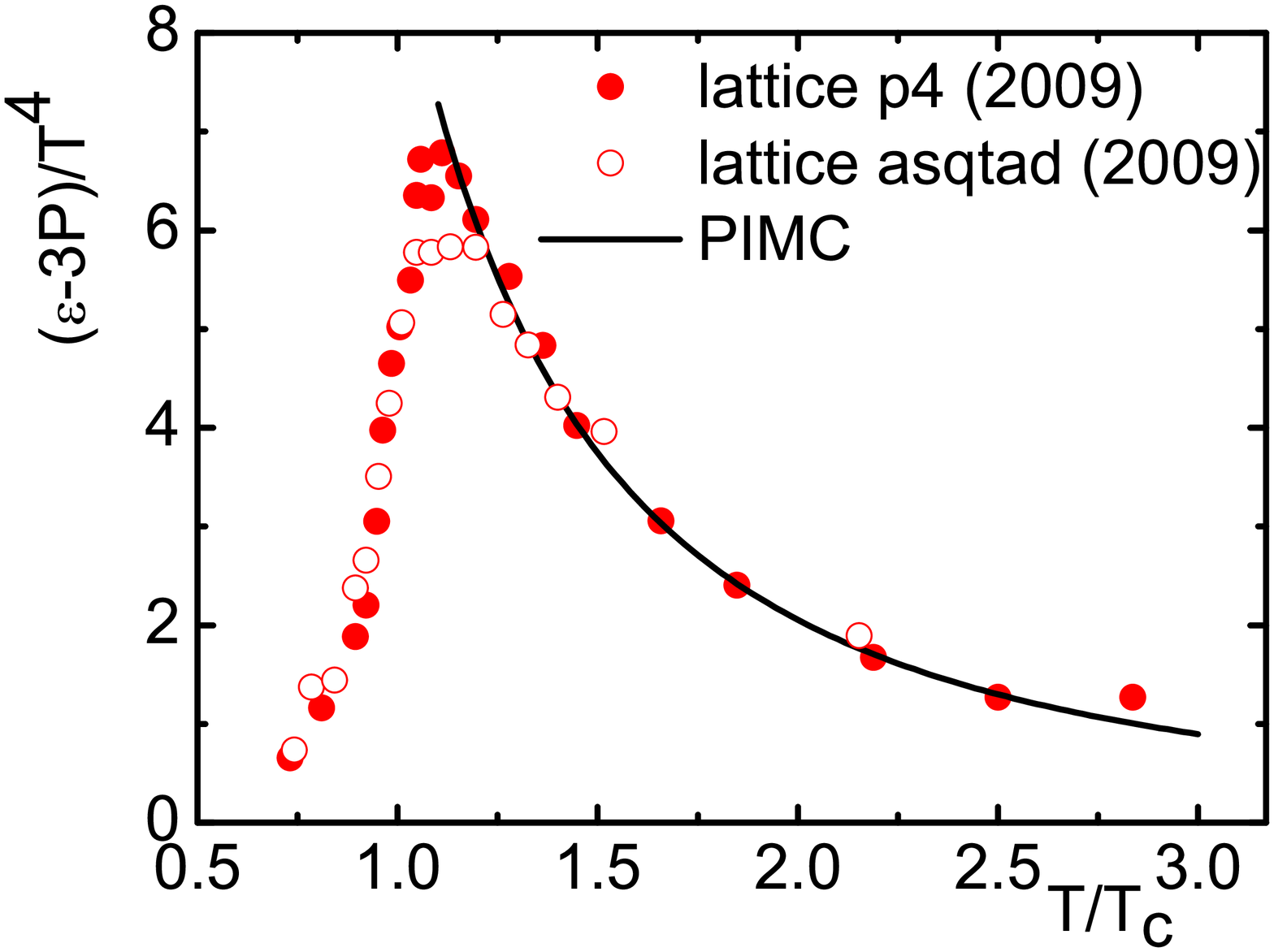}
\caption{
Entropy (left panel) and trace anomaly (right panel) of the QGP from PIMC
simulations (solid line) compared 
to lattice data of Refs.~\cite{Lattice09,Csikor:2004ik}.
Notation is the same as in the right panel of Fig. \ref{fig:EOS}.
}
\label{fig:EOS1}
\end {figure}

Figure \ref{fig:EOS1} additionally presents the entropy ($S$) and
trace anomaly ($\varepsilon-3P$)  of the QGP computed in the PIMC
method. These quantities are calculated accordingly to Eqs.
(\ref{p_gen}). In order to avoid the numeric noise, the derivative of
a smooth interpolation between the 
PIMC points (see the right panel of Fig. \ref{fig:EOS}) was taken.
These results are  compared 
to lattice data of Refs.~\cite{Lattice09,Csikor:2004ik}. It is not
surprising that agreement with the lattice data is also good, since it
is a direct consequence of the good reproduction of the pressure.

Details of our path integral Monte-Carlo simulations have been discussed elsewhere in a variety
of papers and review articles, see,
e.g. \cite{rinton} and references therein. The main idea of the simulations consists in constructing
a Markov process of configurations which differ by the particle coordinates.
Additionally to the case of electrodynamic
plasmas, here  we  randomly sample, according to the group measure, the color variables $Q$ of all particles until convergence is achieved.
We use a cubic simulation box with periodic boundary conditions. The number of particles was
taken as $N=N_q+N_{ \bar{q}}+N_g=40+40+40=120$, and the number of beads, $n=20$.
Calculation of the equation of state (right panel of Fig.~\ref{fig:EOS})
was used to optimize the parameters of the model in order to
proceed to predictions of other properties concerning the internal structure
and in the future also non-equilibrium dynamics
of the QGP.

\begin{figure}[htb]
\vspace{0cm} \hspace{0.0cm}
\includegraphics[width=7.9cm,clip=true]{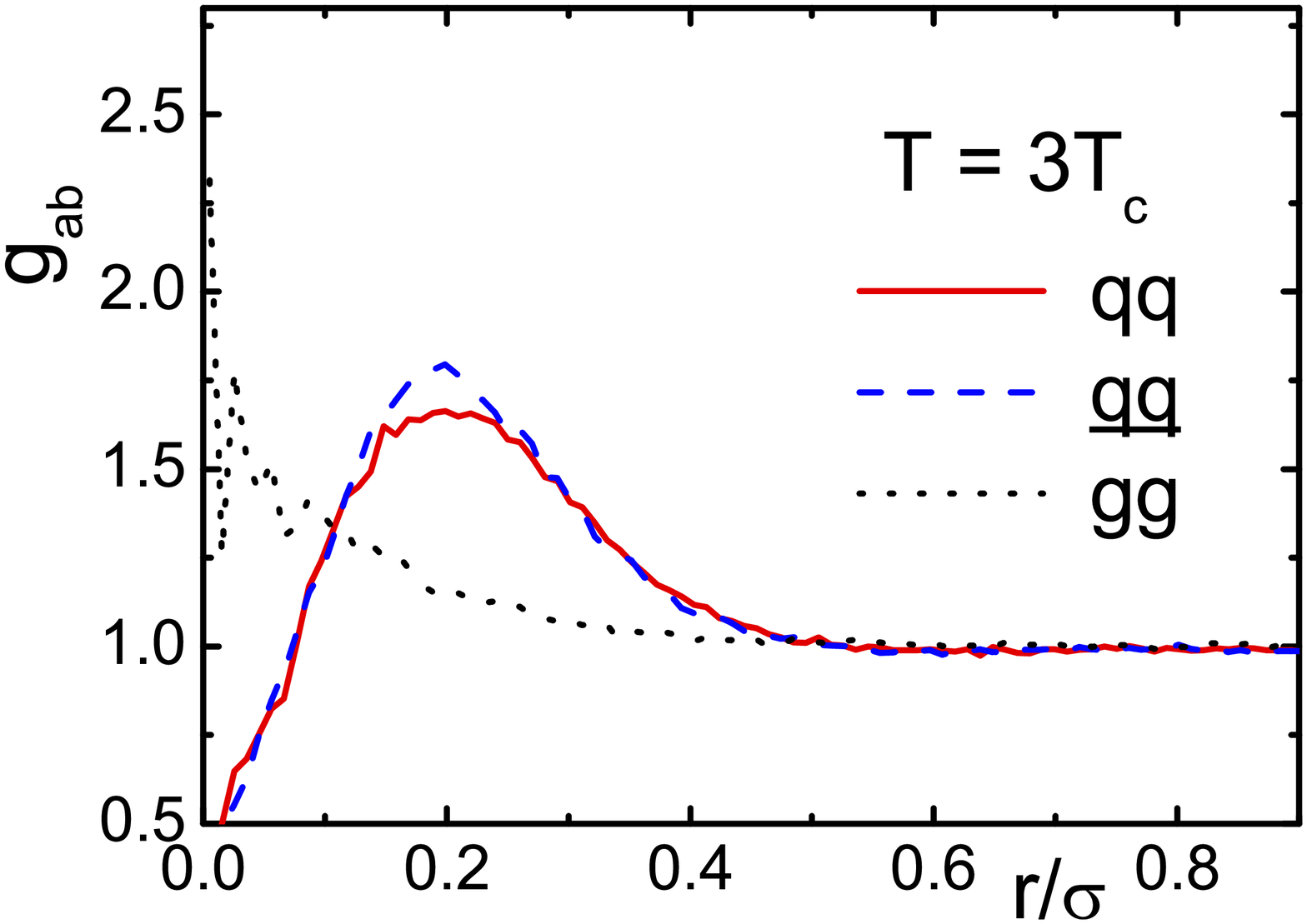}
\includegraphics[width=7.9cm,clip=true]{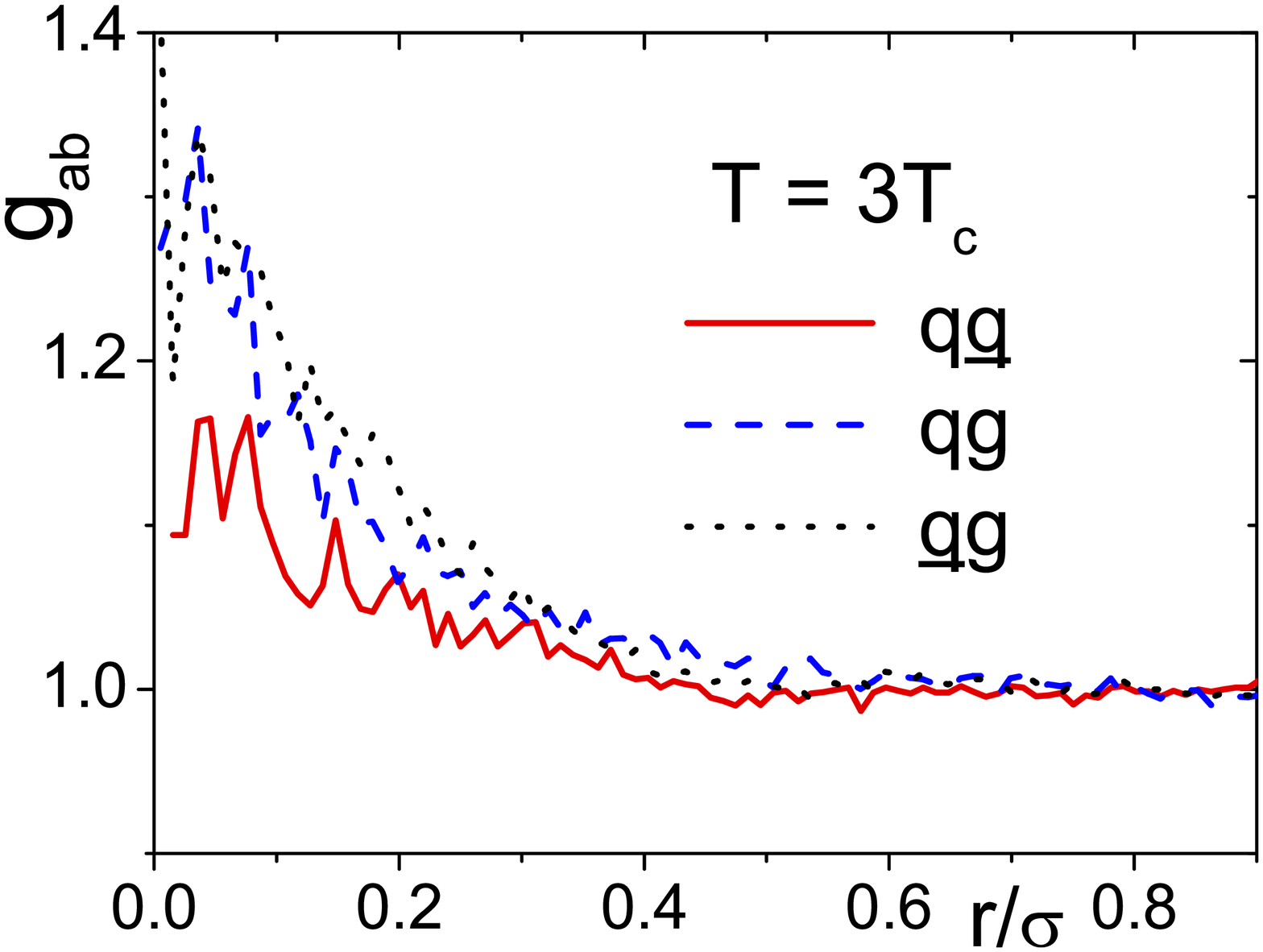}
\includegraphics[width=7.9cm,clip=true]{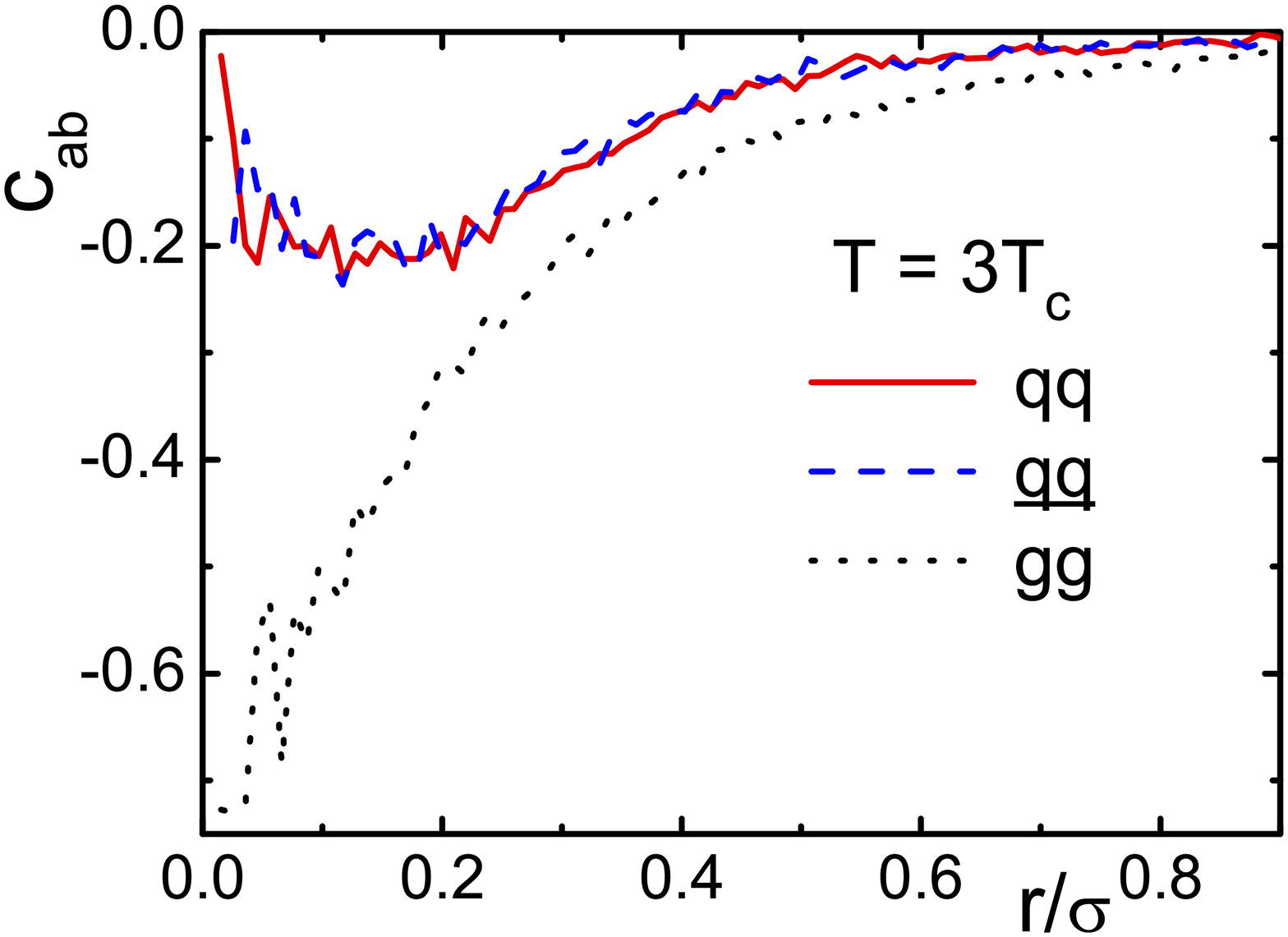}
\includegraphics[width=7.9cm,clip=true]{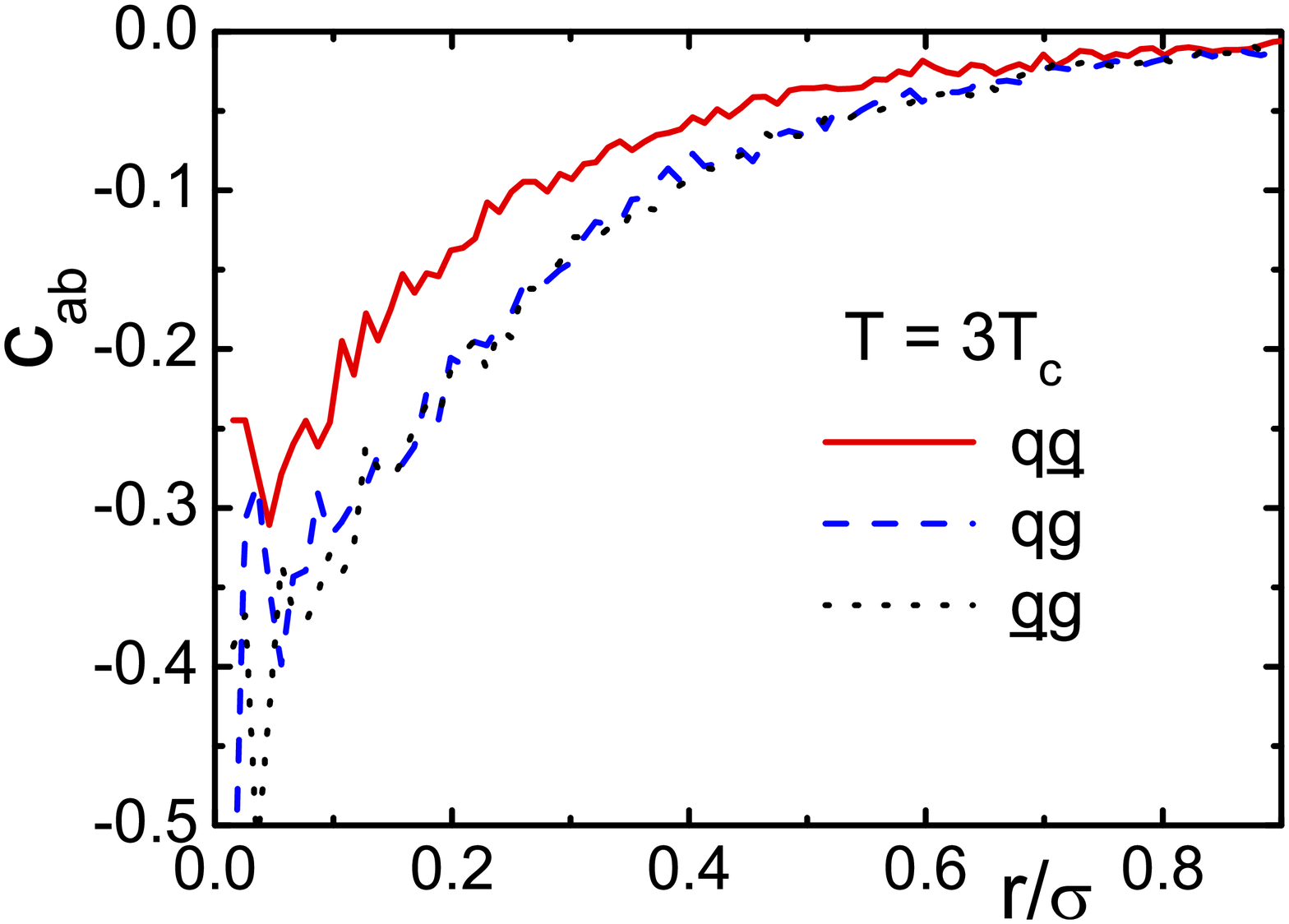}
\caption{Pair distribution functions (upper panels) and color pair distribution
functions (lower panels) of identical (left panels) and different (right panels) quasi-particles  at
temperature $T/T_c=3$.
The distance is measured in units of 
$\sigma = 1/T_c = 1.1$ fm.
}
\label{fig:PDFC}
\end{figure}
Let us now consider the spatial arrangement of the quasi-particles in the QGP by studying the
 pair distribution functions (PDF's) $g_{ab}(r)$. They give the probability density
to find a pair of particles of types $a$ and $b$ at a certain distance $r$ from each other and are
 defined as
\begin{eqnarray}\label{g-def}
g_{ab}(R_1-R_2)=
\frac{1}{Z N_q!N_{ \bar{q}}!N_g!}
\sum_{\sigma}\int
dr dQ\,\delta(R_1-r^a_1)\delta(R_2-r^b_2)\rho(r,Q, \sigma ;\beta).
\end{eqnarray}
The PDF's depend only on the difference of coordinates because of the translational invariance of the system.
In a non-interacting classical system,
$g_{ab}(r)\equiv 1$, whereas interactions and  quantum statistics result in
a re-distribution of the particles.
Results for the PDF's at temperature $T/T_c=3$ are shown in 
the top panels of Fig.~\ref{fig:PDFC}.
%

At large distances, $r/\sigma \ge 0.5$,
all PDF's of identical particles (top left panel of Fig. \ref{fig:PDFC})
coincide. 
A drastic difference in the behavior of the
PDF's of  quarks and gluons (the anti-quark PDF is identical to the quark PDF) occurs
at small distances. Here
the gluon PDF increases
monotonically when the distance goes to zero, while
the PDF of quarks (and antiquarks) exhibits a broad maximum.
This difference is the effect of the quantum statistics.
In the present conditions, the thermal wavelength $\lambda$  approximately equals
$0.37\sigma$, i.e. the difference starts to appear at distances of the order of $\lambda$.
The enhanced population of
low distance states of gluons is due to Bose statistics and the color-Coulomb attraction.
In contrast, the depletion of the small
distance range for quarks is a consequence of the Pauli principle.
In an ideal Fermi gas $g(r)$ equals zero for particles with the
same spin projections and colors, while for particles with different colors and/or
opposite spins the PDF equals unity in the limit $r\to 0$. As a consequence,
the spin and color averaged PDF approaches $0.5$.
Such a low-distance behavior is also observed in a nonideal dense
astrophysical electron-ion plasma and in a nonideal  electron-hole plasmas in semiconductors \cite{bonitz_jpa06,filinov_pre07}.
The depletion of the probability of quasi-particles at small distances results in its enhancement at intermediate
 distances. This is the reason for the corresponding PDF maxima. 
Oscillations of the PDF at very small distances of order
$r\le 0.02 \sigma$ are related to Monte Carlo statistical error, as probability of  quasi-particles being at short distances quickly decreases.

At small distances, $r\le 0.3\sigma$, a strong increase is observed
in all PDF's of particles of different type (top right panel of Fig. \ref{fig:PDFC}),
which resembles the behavior of the gluon-gluon PDF.
This increase
is a clear manifestation of an effective pair attraction
of quarks and antiquarks as well as quarks (antiquarks) and gluons.
This attraction suggests that the color vectors of nearest
neighbors
of any type are anti-parallel.
%
%
If this explanation is correct can be verified by means of
 {\em color pair distribution functions} (CPDF) defined as
\begin{eqnarray}\label{c-def}
c_{ab}(R_1-R_2)=
\frac{1}{Z N_q!N_{ \bar{q}}!N_g!}
\sum_{\sigma}\int
dr dQ\,\langle Q^a_1|Q^b_2 \rangle \delta(R_1-r^a_1)\delta(R_2-r^b_2)\rho(r,Q, \sigma ;\beta),
\end{eqnarray}
%
which are shown in the lower panels of Fig.  \ref{fig:PDFC}.
All CPDF's 
turn out to be negative at small distances, indicating anti-parallel orientation of the color vectors
of neighboring quasi-particles, which
complies with attraction seen in the functions $g_{ab}$ for $a\ne b$.
The minimum of $c_{qq}$ close to $r=0.2\sigma$ corresponds to the maximum observed in $g_{qq}$.
The deep minimum in the
gluon CPDF
at small distances results from the Bose statistics and 
complies with the high maximum of the gluon PDF $g_{gg}$.
%
\begin{figure}[htb]
\vspace{0cm} \hspace{0.0cm}
\includegraphics[width=7.9cm,clip=true]{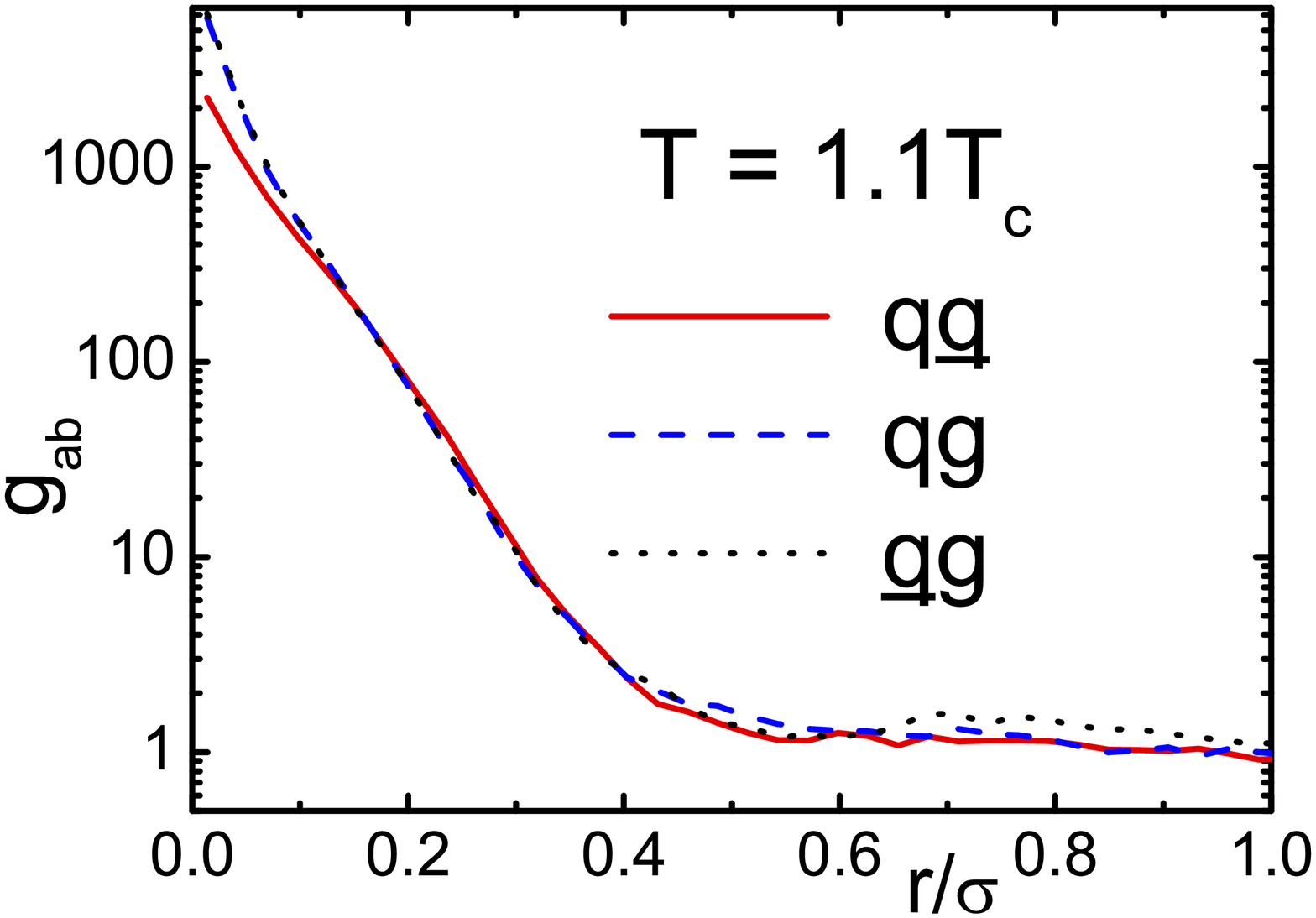}
\includegraphics[width=7.9cm,clip=true]{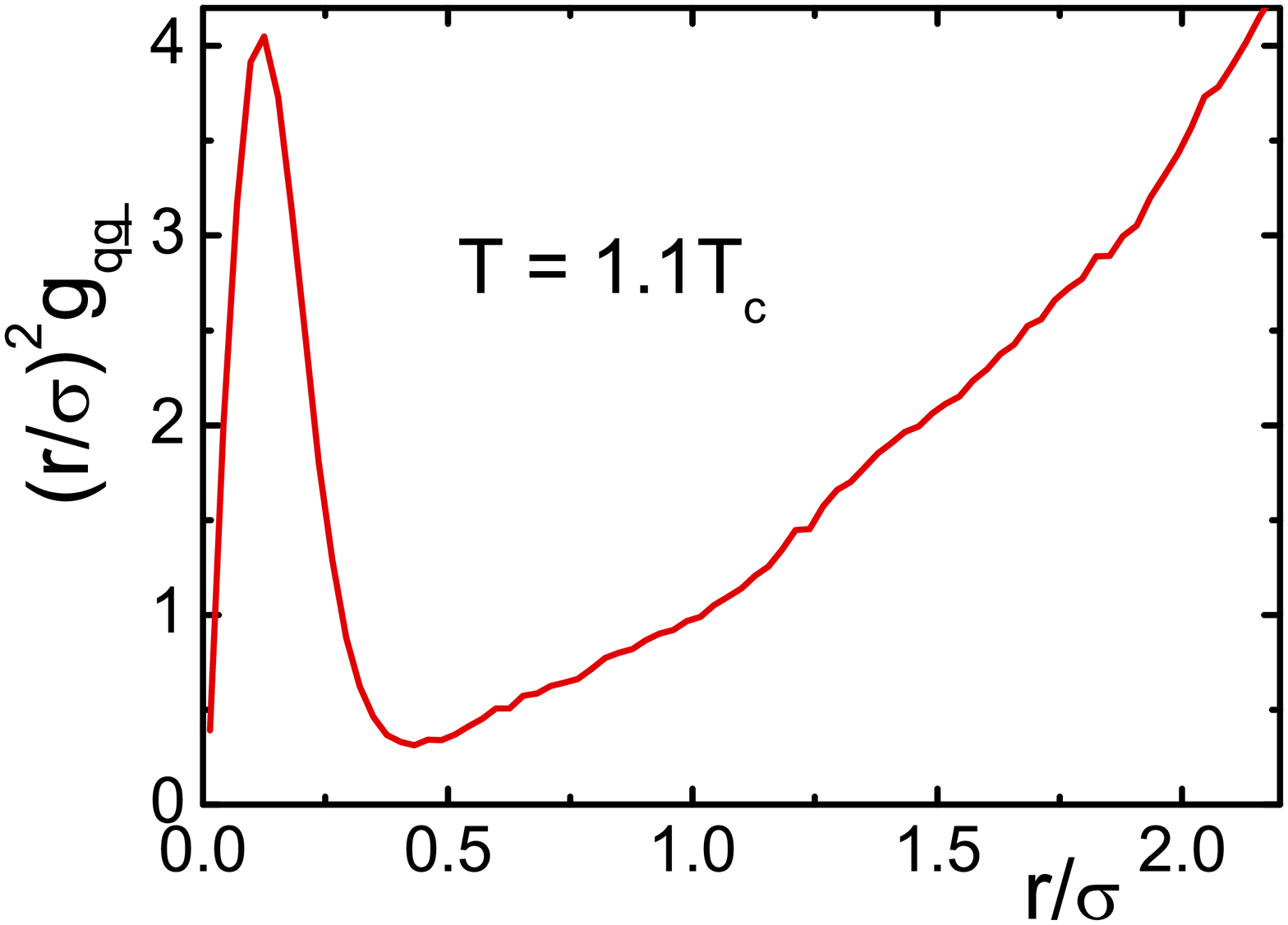}
\includegraphics[width=7.9cm,clip=true]{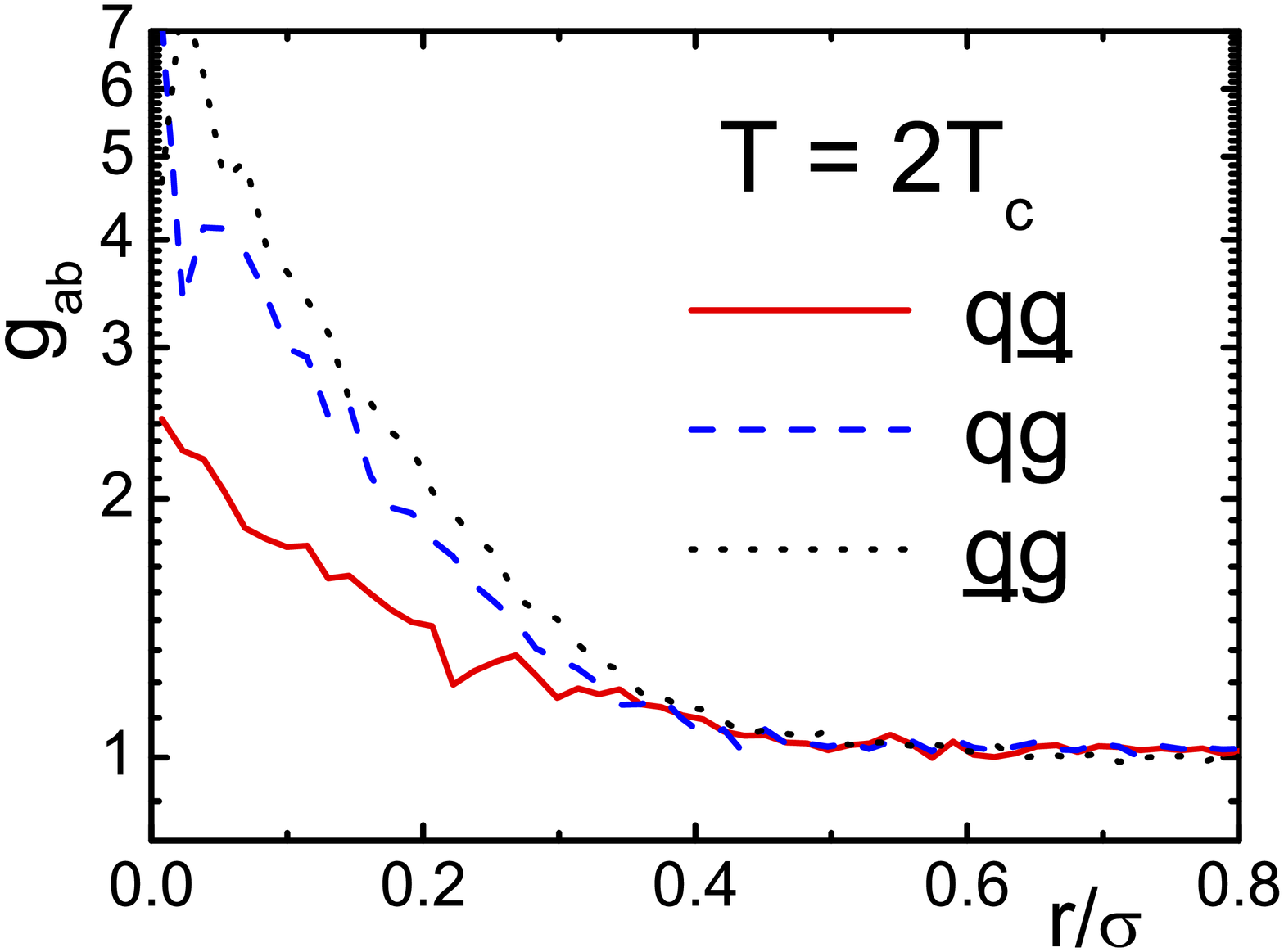}
\includegraphics[width=7.9cm,clip=true]{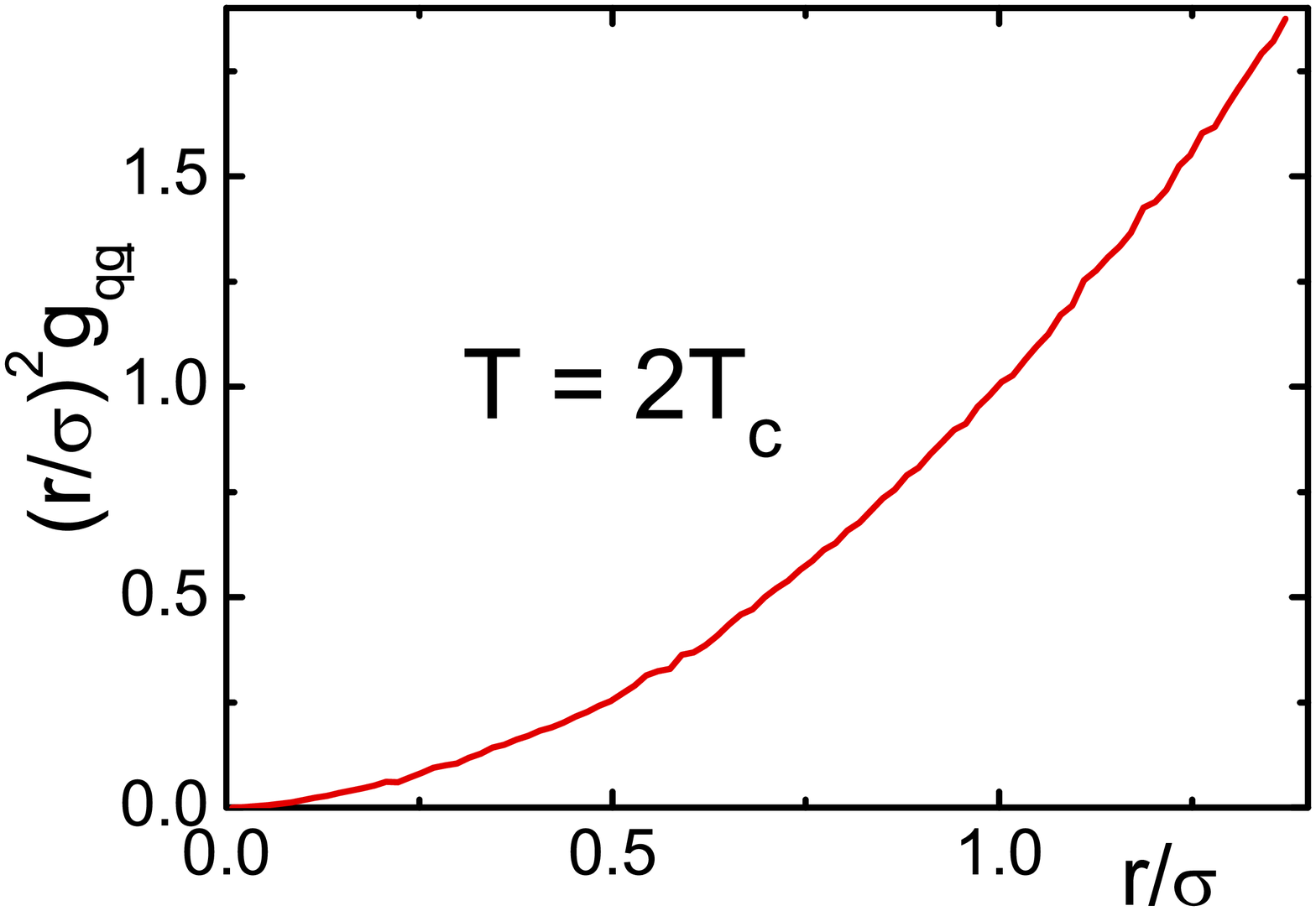}
\includegraphics[width=7.9cm,clip=true]{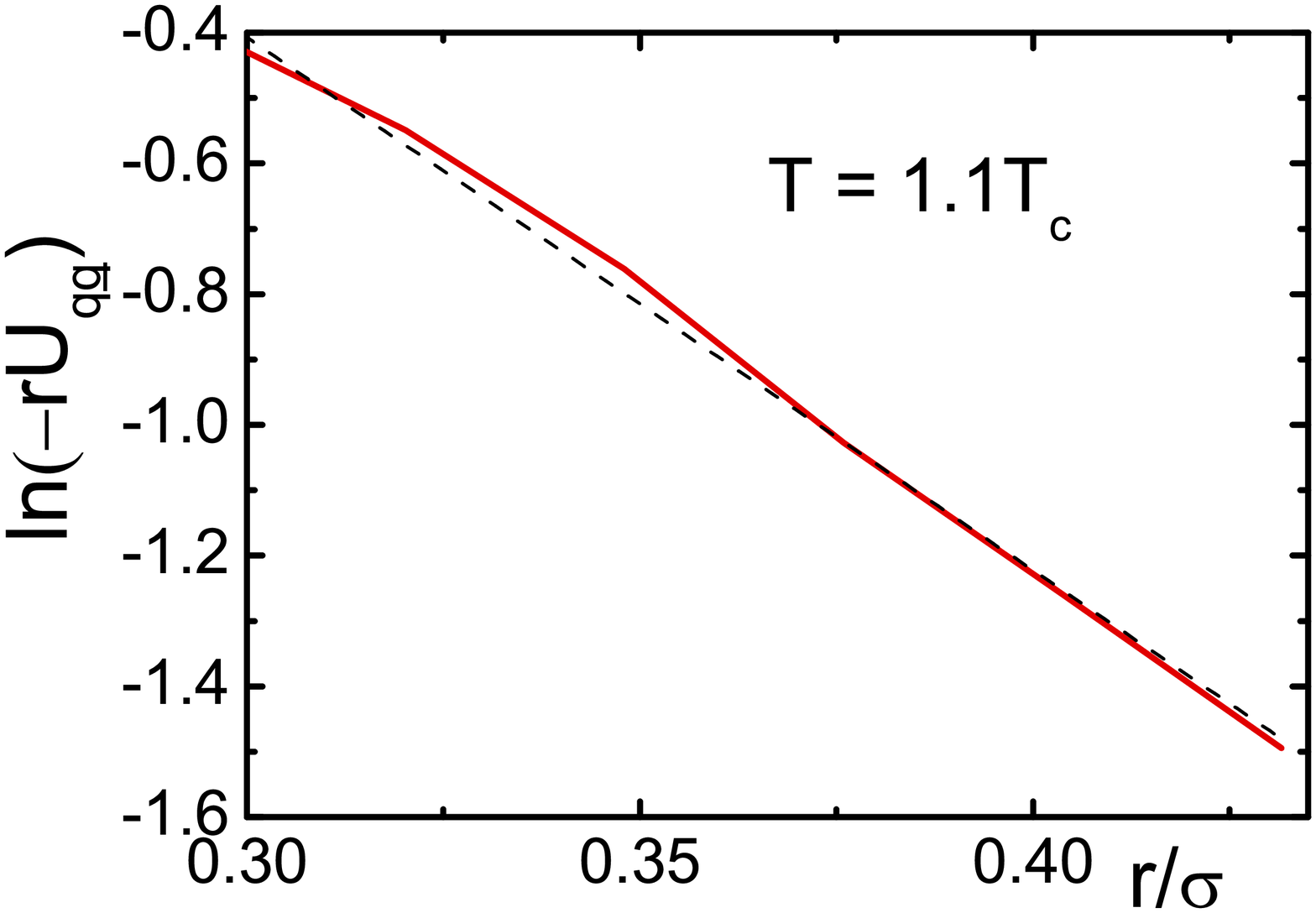}
\includegraphics[width=7.9cm,clip=true]{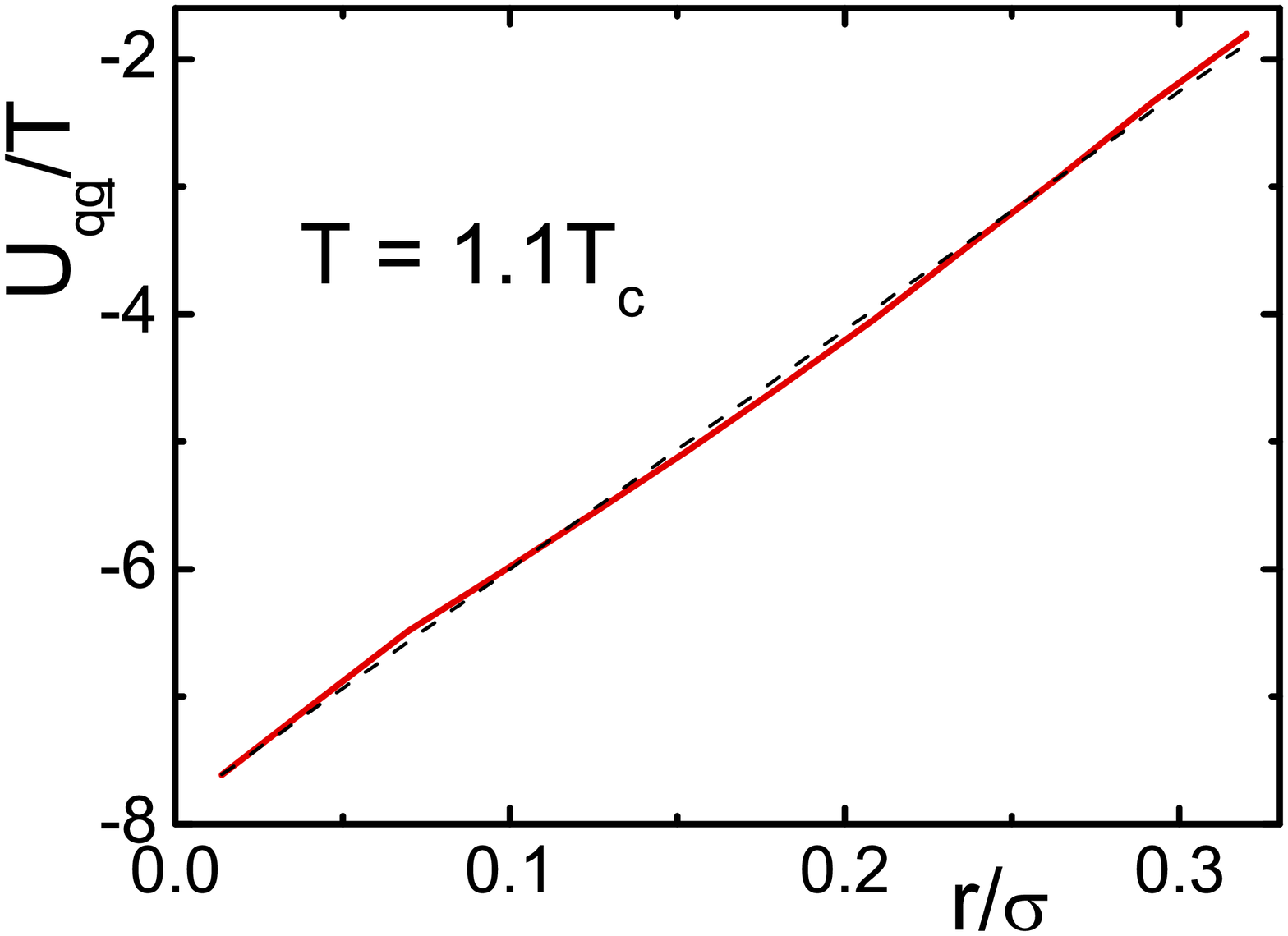}
\caption{
Top and middle panels:
Pair distribution functions at two different temperatures $T$
(left panels) and quark-antiquark PDF multiplied by
distance squared (right panels). The distance is measured in units of 
$\sigma = 1/T_c = 1.1$ fm.
Bottom panels: potential of average force, defined according to Eq.
(\ref{PAF}), 
at small distances for $T=1.1 T_c$ (right panel)
and logarithm of the product $r$ and averaged potential at large distances (left panel). Dashed lines are
linear fits to the calculated (solid) lines. 
}
\label{fig:PDBLSC}
\end{figure}
%

Thus, at $T/T_c=3$
we observe signs of a spatial ordering,
cf. the peak of the quark PDF around $r/\sigma=0.1-0.2$,
which may be interpreted as emergence of liquid-like behavior of the QGP.
Much more pronounced is the short-range structure of nearest neighbors.
The QGP lowers its total energy by minimizing the color
Coulomb interaction energy via a spontaneous ``anti-ferromagnetic'' ordering of color vectors. 
This gives rise to a clustering
of quarks, antiquarks and gluons.
To verify the relevance of these trends, a more refined spin-resolved analysis of
 the PDF's and CPDF's is necessary, together with simulations
in a broader range of temperatures which are presently in progress.

PDF's allow us  also to
do a lot of physical conclusions rather than only to analyze the spatial structure of the QGP.
Fig. \ref{fig:PDBLSC} presents PDF's of the identical particles
for two temperatures $T=1.1T_c$ and $T=2T_c$ (left upper and middle panels).
These PDF's can be formed either by correlated scattering states  or by bound states of quasi-particles,
depending on the relative fractions of these states.
In a strict sense, however,
there is no clear subdivision into bound and free ``components''
due to the mutual overlap of the quasi-particle clouds. In addition, there exists no
rigorous criterion for a bound state at high densities due to the strong effect of the surrounding
plasma. 
Nevertheless, a rough estimate
of the fraction of quasi-particle bound states can be obtained
by  the following reasonings. The product $r^2 g_{ab}(r)$
has the meaning of a probability to find
a pair of quasiparticles at a distance $r$ from each other.
On the other hand, the corresponding quantum mechanical
probability is the product of $r^2$ and the two-particle Slater sum
\begin{equation}
\label{slsm}
\Sigma_{ab}=8\pi
^{3/2}\lambda_{ab}^{3}\sum_{\alpha}|\Psi_{\alpha}(r)|^{2}\exp(-\beta
E_{\alpha})
 = \Sigma_{ab}^{d}+\Sigma_{ab}^{c},
\end{equation}
where $E_{\alpha}$ and $\Psi_{\alpha}(r)$ are
the energy (without center of mass energy) and the wave
function of a quasi-particle pair.
, respectively, and $\lambda_{ab}=\sqrt{2\pi \beta (m_{a}+m_{b})/(m_{a}m_{b})}$.
$\Sigma_{ab}$ is, in essence,
the diagonal part of the corresponding density matrix.
The summation runs over all 
states $\alpha$ of
the discrete ($\Sigma_{ab}^{d}$) and
continuous ($\Sigma_{ab}^{c}$)
spectrum.
The discrete-spectrum contribution
\begin{equation}
\Sigma_{ab}^{d}=8\pi^{3/2}\lambda_{ab}^{3}\sum_{E_{\alpha}=
E_{0}}^{E'}|\Psi_{\alpha}(r)|^{2}\exp(-\beta E_{\alpha})
\end{equation}
originates from the populated states between the ground one with the energy $E_{0}$ and
and some state with the energy $E'$.
For low densities it is
reasonable to choose
$E' \approx - T$  \cite{zamalin}.

At temperatures smaller than the binding energy and
distances smaller than or of the order of several bound state radii, the main
contribution to the Slater  sum comes from bound
states, while at larger distances free (scattering) states give a
dominant contribution. 
In the electromagnetic plasma it was found that
the product
$r^2\Sigma_{ab}^d$ is sharply peaked at distances around the Bohr
radius in this case. 
Similarly, at low temperature, $r^2g_{q\bar{q}}(r)$ forms a pronounced
maximum near 
$r=0.2$ fm which can be interpreted as the radius of a bound
$q\bar{q}$ pair (see right upper panel of Fig. \ref{fig:PDBLSC}).   
Thus, our calculations support the existence of bound states of medium-modified
(massive) quarks and gluons at moderate temperatures, i.e. just above $T_c$,
proposed in Ref. \cite{Yukalov97} and later in Refs. \cite{Shuryak03,Brown05}
based on results from lattice QCD calculations
of spectral functions \cite{Asakawa01,Karsch03}.
Integration of $c_{q\bar{q}}$ from zero up to $0.3$ fm shows that
modulus scalar product 
$|\langle Q_q+Q_{\bar{q}}|Q_q+Q_{\bar{q}}\rangle |$ is much smaller that $|\langle Q_q|Q_q\rangle|$
and $|\langle Q_{\bar{q}}|Q_{\bar{q}}\rangle|$
that means that these bound states are colorless.
With the temperature rise these bound states dissolve much faster than it was assumed in
\cite{Shuryak03,Brown05}, which complies with the analysis of Ref. \cite{Koch05}.
Indeed, at the temperature of $T= 2 T_c$ the bound states completely disappear
(see right middle panel of Fig. \ref{fig:PDBLSC})
and $r^2 g_{ab}(r)$ approaches the behaviour which corresponds 
to the plasma without bound states.

Interesting observations can be done from the analysis of
the potential of average force (PAF) defined as the logarithm of  the related PDF,
\begin{equation}
\label{PAF}
U_{ab}(r,T) = -T \ln g_{ab} (r,T)
\end{equation}
(lower panels of Fig.~\ref{fig:PDBLSC}).
This definition is motivated by the PDF virial expansion in terms of bare potential (like color Kelbg potential).
Near the QGP phase transition the PAF (right bottom panel)
is a linear function at distances smaller than the bound state radius.
This suggests that the bound states are bound by a string-like forces.
At larger distances the PAF (left bottom panel)
can be very well approximated
by an exponentially screened Coulomb potential (Yukawa-type potential)
like that in the electromagnetic plasma.
Potential well of the linear part of average force potential is of order $1.5$ GeV at $T=1.1 T_c$, while for $T = 2 T_c$ it is only of order $0.2$ Gev.
Straight dashed lines in the lower panels of Fig.~\ref{fig:PDBLSC} are the least-square
liner (left panel) and Yukawa-type (right panel) approximations to the the effective potential.

\section{Conclusion}\label{s:discussion}

Quantum Monte Carlo simulations 
based on the quasi-particle picture of the QGP
are able to reproduce the lattice equation of state (even near the critical temperature) and also
yield valuable insight into the internal structure of the QGP.
Our results indicate that the QGP reveals liquid-like (rather than gas-like) properties even at the
highest considered temperature of $3T_c$. At temperatures just above $T_c$ we have found that
bound quark-antiquark states still survive. These states are bound by effective string-like forces.
Quantum effects turned out to be of prime importance in these  simulations.



Our analysis is still too simplified and incomplete. It is still confined only to the case of zero
baryon chemical potential. The input of the model also requires refinement.
Work on these problems is in progress. 

However, the PIMC method is not able to yield dynamical and transport properties of the QGP.
One way to achieve this is to develop semiclassical color molecular dynamics simulations. In contrast to previous MD simulations \cite{shuryak1}, where quantum effects were included phenomenologically via a short range potential,
we have developed a more rigorous approach to study the dynamical and transport properties of strongly coupled 
quark-gluon systems
which is based on the combination of the Feynman and Wigner formulation of quantum dynamics. 
The basic ideas of this approach for  the electron-ion plasmas have been published in \cite{afilinov_jpa03,afilinov_pre04,rinton}. 
In particular, this approach allows us to deduce the viscosity of
the QGP.  Work on these problems is also in progress.  

%

We acknowledge stimulating discussions with D.~Blaschke, M.~Bleicher, R. Bock, B.~Friman,
C. Ewerz, D.~Rischke, and H.~Stoecker.
Y.I. was partially supported by the Deutsche  Forschungsgemeinschaft
(DFG projects 436 RUS 113/558/0-3 and WA 431/8-1),
the RFBR grant 09-02-91331 NNIO\_a, and grant NS-7235.2010.2.


\bibliographystyle {apsrev}

\end{document}